\documentclass[amsmath,amssymb,floatfix,prb,epsf,showpacs,twocolumn,superscriptaddress]{revtex4-2}
\usepackage{standalone}
\usepackage{graphics}
\usepackage{overpic}
\usepackage{pict2e}
\usepackage{amsmath,amsfonts}
\usepackage{bigints}
\usepackage{amscd}
\usepackage{makeidx}
\usepackage{multirow}
\usepackage{mathtools}
\usepackage{amsfonts}
\usepackage[dvipsnames]{xcolor}
\usepackage[caption=false]{subfig}
\usepackage{siunitx}
\usepackage{booktabs}
\usepackage{xcolor}
\usepackage{tikz}
\usetikzlibrary{3d} 
\usetikzlibrary {arrows.meta}
\usepackage{soul}
\setstcolor{red}
\tikzset
{
   axis/.style={thick,-latex},
   yz/.style={canvas is yz plane at x=0},
   xz/.style={canvas is xz plane at y=0},
   xz rotated/.style={rotate around z=\mytheta,xz},
   xy elevated/.style={canvas is xy plane at z={\myradius*cos(\myphi)},scale={sin(\myphi)}},
   sphere/.style={shading=ball,fill opacity=0.3},
   plane/.style={fill=teal,fill opacity=0.3},
   cone/.style={fill=yellow,fill opacity=0.3},
   vector/.style={very thick,-stealth},
   point/.style={shading=ball,ball color=red}
}
\usetikzlibrary{shapes,calc,arrows,through,intersections}
\usetikzlibrary{arrows.meta,calc,shapes,decorations.pathreplacing,decorations.markings, snakes}

\usepackage[compat=1.1.0]{tikz-feynman}
\usepackage{feynmp-auto}
\usepackage{epic,eepic,epsfig}
\usepackage{dcolumn}
\usepackage{bm}
\usepackage{color}
\usepackage{bbm}
\usepackage{physics}
\newcommand{\nn}{\nonumber}
\newcommand{\bea}{\begin{eqnarray}}
\newcommand{\ena}{\end{eqnarray}}
\usepackage{pgfplots}
\pgfplotsset{compat=newest}

\usepackage{xcolor}
\colorlet{linkequation}{blue}
\usepackage[colorlinks]{hyperref}
\usepackage{cleveref}
\usepackage[toc,title,page]{appendix}
\usepackage{ragged2e}

\makeatletter
\long\def\@makecaption#1#2{%
  \par\vskip\abovecaptionskip
  \begingroup
    \small\rmfamily
    \setlength{\parindent}{0pt}%
    \makebox[\linewidth][l]{%
      \begin{minipage}{\linewidth}
        \justifying
        \@make@capt@title{#1}{#2}\par
      \end{minipage}%
    }%
  \endgroup
  \vskip\belowcaptionskip
}
\makeatother

\hypersetup{
    colorlinks=true,
    linkcolor=red,
    filecolor=blue,      
    urlcolor=blue,
    citecolor=green,
    pdftitle={Overleaf Example},
    pdfpagemode=FullScreen,
    }
\makeatother

\newcommand{\yerphiadd}{ A.I. Alikhanyan National Science Laboratory, Yerevan Physics Institute, 2 Alikhanyan brothers str, Yerevan 0036, Armenia}
\newcommand{\umassadd}{Department of Physics, University of Massachusetts, 710 North Pleasant Street, Amherst, MA 01003-9337, USA}

\begin{document}

\title{Nonequilibrium from Equilibrium: Chiral Current-Carrying States in the Spin-1 Babujian-Takhtajan Chain}

\author{Bahar Jafari--Zadeh}
    \affiliation{\umassadd}
       
\author{Chenan Wei}
       \affiliation{\yerphiadd}
         \affiliation{\umassadd}
\author{Hrachya M. Babujian}
    \affiliation{\yerphiadd}

\author{Tigran A. Sedrakyan}
    \affiliation{\umassadd}
  \affiliation{\yerphiadd}
\date{\today}

\begin{abstract}
We study the spin-$1$ Babujian-Takhtajan chain deformed by its third conserved charge $Q_3$. We derive  $Q_3$ and show that it is a dimensionless energy current and that its local density is a dressed scalar-chirality operator rather than bare chirality alone, as is the case for the spin-$1/2$ Heisenberg chain. The deformation $H_\alpha=H+\alpha Q_3$ therefore provides a local, exactly solvable current bias: it leaves the eigenstates of the original Hamiltonian unchanged, but reorders them so that selected high-energy current-carrying states become ground states of the tilted problem. Using the thermodynamic Bethe ansatz and confirming the analytical calculations with DMRG, we find a quantum phase transition at $\alpha_c={J}/(8\pi)$.
For $\alpha<\alpha_c$, the ground-state remains the undeformed Babujian-Takhtajan  phase whose low-energy effective field theory is described by the 
$SU(2)$  Wess-Zumino-Witten (WZW) model at level $k=2$ representing a critical phase characterized by a central charge $c=3/2$ and $\langle Q_3\rangle=0$. For $\alpha>\alpha_c$, a finite rapidity interval forms, and the system enters a gapless chiral current-carrying sector described by a $c=3/2$ CFT. Near the threshold, the free energy starts quadratically as a function of $\alpha-\alpha_c$, while the energy current turn on linearly. 
The scalar chirality 
turns on at the same threshold, showing that the postcritical sector is simultaneously current-carrying and chiral. 
The most immediate experimental routes are composite spin-1 bosons in optical lattices, and programmable qutrit simulators based on trapped ions or superconducting circuits.
\end{abstract}

\date{\today}
\maketitle

\section{Introduction}
\label{sec:I}
Physics of ground states and low-lying excitations in quantum many-body systems can be successfully analyzed within field theory, bosonization,  Bethe ansatz, density-matrix renormalization group (DMRG), Monte Carlo simulations, and tensor-network approaches. However, many physically important sectors lie much higher in the spectrum. This is particularly the case for states carrying a finite current, a finite chirality, or, more generally, definite values of conserved quantities other than the energy. In a generic interacting system, such states are finite-energy-density eigenstates, thereby raising questions of thermalization and of the eigenstate thermalization hypothesis (ETH): in the absence of additional structure, a typical highly excited eigenstate is expected to look locally thermal~\cite{Deutsch-1991,Srednicki-1994,Rigol-2008,DAlessio-2016,Gogolin-Eisert-2016,Mori-2018}.

The central idea of this paper is that this obstacle can be avoided exactly whenever suitable conserved operators are known explicitly. If a Hamiltonian $H$ commutes with a charge $Q$, introducing a term proportional to $Q$ leaves the eigenstates unchanged and only reorders the spectrum. A conserved-charge deformation~\cite{Tsvelik,Frahm,MkhitaryanSedrakyan,VagharshTigran} can therefore turn a highly excited sector of the original model into the ground state of a new, still local Hamiltonian. This construction enables the use of ground-state methods to analyze such a sector, while allowing the measurement of observables in it that are not themselves conserved.

Closely related constructions in spin-$\tfrac12$ chains have already revealed current-carrying and chirality-carrying critical sectors with broken parity and time-reversal symmetry, a linear onset of chirality, and characteristic finite-size scaling~\cite{Wei-Mkh-Sed-2024,SedrakyanPangWeiWang-2025}. More broadly, time-reversal-breaking criticalities and symmetry-enriched critical points exhibit universal scaling and Lifshitz-type reconstructions of the low-energy spectrum~\cite{WangSedrakyan-2020,WangSedrakyan-2022}, while chiral order is also central in higher-dimensional settings, ranging from helical spin liquids on frustrated magnets to chiral vortex-line liquids, and moat-band chiral spin liquids~\cite{SedrakyanMoessnerKamenev-2020,CSL1,CSL2,CSL3,CSL4,CSL5,JafariZadehWeiSedrakyan-2025}. 

The spin-$1$ Babujian-Takhtajan (BT) chain \cite{Babuj-1982,Takh-1982,Babuj-1983,Zam-Fat-1980,Fad-Takh-1987,Fad-96,SB-2013} provides a particularly clear framework in which this general strategy can be developed exactly. 
The Hamiltonian is given by
\begin{equation}
H = \frac{J}{4}\sum_{n=1}^{N}\left( \mathbf{S}_n \cdot \mathbf{S}_{n+1} - (\mathbf{S}_n \cdot \mathbf{S}_{n+1})^2 \right)
\;,
\label{eq:BT-H}
\end{equation}
where $\mathbf{S}_n$ is a spin-$1$ operator at site $n$, and $J>0$ sets the antiferromagnetic exchange scale. Eq.~\eqref{eq:BT-H} can be regarded as the bilinear-biquadratic spin-$1$ Heisenberg chain at its integrable point and in the thermodynamic limit, it corresponds to the $SU(2)_2$ WZW conformal field theory with a central charge of $c=3/2$. A $q$-deformation of spin-$1$ BT chain was studied in Ref.~\cite{Quella-2021}. It is convenient to separate the overall scale $J/4$ from the bond operator,
\begin{equation}
h_n=\mathbf{S}_n\cdot\mathbf{S}_{n+1}-(\mathbf{S}_n\cdot\mathbf{S}_{n+1})^2
\;,
\qquad
H=\frac{J}{4}\sum_{n=1}^{N}h_n
\;.
\label{eq:hn-intro}
\end{equation}
The BT chain is ideal for the present purpose for two reasons. First, it is interacting and genuinely spin-$1$, so the current-carrying sector is not a free-fermion artifact. Second, its conserved charges can be derived explicitly, which allows the nonequilibrium sector selected by the tilt to be characterized analytically and numerically on equal footing.

The general construction can be stated without using any special property of the BT model. Suppose that, in addition to $H$, one knows mutually commuting operators $\{Q_a\}$ such that
\[
[H,Q_a]=0, \qquad [Q_a,Q_b]=0 \;.
\]
These operators are constants of motion. One may then introduce a tilted Hamiltonian
\begin{equation}
H_{\{\lambda_a\}} = H + \sum_a \lambda_a Q_a\;,
\label{eq:tilted-general}
\end{equation}
where the fields $\lambda_a$ are conjugate to the conserved quantities. If $|\nu\rangle$ is a simultaneous eigenstate of $H$ and all $Q_a$,
\begin{equation}
H|\nu\rangle = E_\nu |\nu\rangle, \qquad Q_a|\nu\rangle = q_{\nu,a} |\nu\rangle\;,
\label{eq:common-eigen}
\end{equation}
then
\begin{equation}
H_{\{\lambda_a\}}|\nu\rangle
=
\left(E_\nu+\sum_a \lambda_a q_{\nu,a}\right)|\nu\rangle
\;.
\label{eq:tilted-eigen}
\end{equation}
The tilted Hamiltonian therefore has exactly the same eigenstates as the original problem, but it reorders them according to a linear functional in the conserved charges. Its ground state selects the eigenstate of $H$ that minimizes $E_\nu+\sum_a \lambda_a q_{\nu,a}$. From the perspective of the original Hamiltonian, this selected state is generically a highly excited eigenstate. In this way, an excited-state problem is reformulated as a ground-state problem.

This point of view is powerful for two complementary reasons. First, it opens high-energy sectors to methods that are intrinsically strongest at low energies. Second, it cleanly separates the operator that selects the sector from the operators that diagnose its physical content. In practice, this separation matters: the observable that labels the sector need not be the most transparent order parameter for the phase realized in that sector.

Integrable systems are the natural laboratory for this strategy because they possess extensive hierarchies of local or quasilocal conserved quantities~\cite{Vidmar-Rigol-2016,Essler-Fagotti-2016,Ilievski-2015,Prosen-Ilievski-2013,Prosen-2014,Zotos-1997,Calabrese,Pozsgay-2020}. In such models, higher charges often admit a direct physical interpretation as particle currents or energy currents. Once that identification is made, the tilt is no longer an abstract algebraic device: it becomes a local current bias, and the ground state of the tilted Hamiltonian can be interpreted as a current-carrying eigenstate of the original model.

At the same time, the logic is broader than integrability. What is essential is not Bethe ansatz itself, but rather explicit knowledge of conserved operators. This happens in several important nonintegrable settings. In many-body-localized phases, one has quasilocal integrals of motion (``$l$-bits'') that organize the spectrum and obstruct conventional thermalization~\cite{Serbyn-Papic-Abanin-2013,Huse-Nandkishore-Oganesyan-2014,Imbrie-Ros-Scardicchio-2017,Nandkishore-Huse-2015}. In lattice gauge theories and constrained gauge-matter models, Gauss-law generators partition the Hilbert space into dynamically disconnected sectors and can produce disorder-free localization or slow dynamics~\cite{Kogut-Susskind-1975,Brenes-2018,Papaefstathiou-2020}. In Hilbert-space-fragmented systems, exact conservation of charge, dipole moment, or more intricate invariants can split the Hilbert space into many Krylov sectors with anomalous transport and restricted thermalization~\cite{Sala-2020,Morningstar-2020,Singh-2021,Lee-Pal-Changlani-2021,Khudorozhkov-2022,Lydzba-2024-PRL}. A related class of examples is provided by models with exact atypical eigenstates, exact many-body scars, $\eta$-pairing towers, or other analytically controlled embedded sectors~\cite{Yang-1989,Shiraishi-Mori-2017,Turner-2018,Moudgalya-2018,Lin-Motrunich-2019,Pakrouski-2020,Moudgalya-2020,Mark-Motrunich-2020,Schecter-Iadecola-2019,Serbyn-Abanin-Papic-2021,Chiba-Yoneta-2024}.

Recent work~\cite{Frey-2024,Aditya-2024,Lydzba-Vidmar-2024,Kliczkowski-2024,Prelovsek-Nandy-Mierzejewski-2024,Nandy-2024,Pawlowski-2024} has sharpened this perspective further by developing diagnostics of constrained thermalization and nonergodicity based on the block inverse participation ratio, subspace-restricted thermalization, weak ETH, fading ergodicity, emergent dipole conservation, subdiffusion in Stark chains, and long-lived prethermal regimes. Closely related studies explored interaction effects in tilted chains, freezing transitions in dipole-conserving systems, fragmentation-tuned dynamics, and correlation-matrix diagnostics of scarred states~\cite{Krajewski-2025,Ganguli-2025,ClassenHowes-2025,Swietek-2025,Yao-Zhang-2025,Lisiecki-2025}. Complementary to these developments, rigorous no-go results for the mixed-field Ising chain and for higher-dimensional quantum Ising models identify situations in which no analogous local-charge construction exists and conventional thermalization is therefore expected to be robust~\cite{Chiba-2024,Chiba-2025}. Taken together, these developments make clear that explicit constants of motion are not a technical curiosity: they determine which parts of the spectrum can be isolated cleanly and which nonequilibrium sectors can be reformulated as ground-state problems.

The BT chain is particularly attractive because its conserved hierarchy can be derived in closed form. The local charges are generated recursively by the boost operator~\cite{Fad-Takh-1987,SB-2013}
\begin{equation}
\mathcal{B}=\sum_{n=1}^{N} n\, h_n ,
\qquad
Q_{m+1}=i[\mathcal{B},Q_m],
\qquad
Q_2=\sum_{n=1}^{N} h_n
\;.
\label{eq:boost-intro}
\end{equation}
Here, $Q_2$ is the dimensionless Hamiltonian, and $Q_1$ is the Noether current corresponding to a global rotational symmetry, namely, total magnetization. The first nontrivial higher charge is
\begin{equation}
Q_3=i[\mathcal{B},Q_2]
=-i\sum_{n=1}^{N}[h_n,h_{n+1}]
\;,
\label{eq:Q3-intro-def}
\end{equation}
where the second equality follows because bond operators with disjoint support commute. In Sec.~\ref{sec:II}, we derive $Q_3$ and show that it has a direct physical interpretation: up to the overall factor $(J/4)^2$ inherited from Eq.~\eqref{eq:BT-H}, it is the energy-current operator of the spin-1 BT chain. At the same time, its local density is not the bare scalar chirality alone. It contains the total scalar chirality operator, $\sum_{i=1}^N\mathbf{S}_i\cdot(\mathbf{S}_{i+1}\times\mathbf{S}_{i+2})$, but it also contains an additional sum of local terms generated by the biquadratic interaction. In the BT model, the third conserved quantity is therefore best understood as a \emph{dressed chiral energy current}.

This identification leads to the deformation studied in the rest of the paper,
\begin{equation}
H_\alpha = H+\alpha Q_3
\;.
\label{eq:Halpha-intro}
\end{equation}
Because $[H,Q_3]=0$, the Hamiltonians $H$ and $H_\alpha$ share all eigenstates. The ground state of $H_\alpha$ is therefore an eigenstate of the original BT Hamiltonian with a definite value of the conserved current quantum number. By the Hellmann--Feynman theorem,
\begin{equation}
\frac{1}{N}\langle Q_3\rangle = \frac{d f}{d\alpha}
\;,
\label{eq:q3-derivative-intro}
\end{equation}
where $f$ is the ground-state energy density of the tilted Hamiltonian. The natural order parameter characterizing the handedness of the selected state is the scalar chirality density
\begin{equation}
\chi \equiv \left\langle \mathbf{S}_i\cdot(\mathbf{S}_{i+1}\times \mathbf{S}_{i+2})\right\rangle
\;.
\label{eq:chi-intro}
\end{equation}
Related staggered anisotropic chirality constructions were studied in Refs.~\cite{Sedrakyan-2001, Ambjorn-2001, Daniel-2004}. An important conceptual point is that $\chi$ and the conserved current density are distinct observables: the tilt selects the sector through $Q_3$, whereas $\chi$ reveals whether that sector is genuinely chiral.

\begin{figure}[t]
\centering
\begin{overpic}[width=\linewidth]{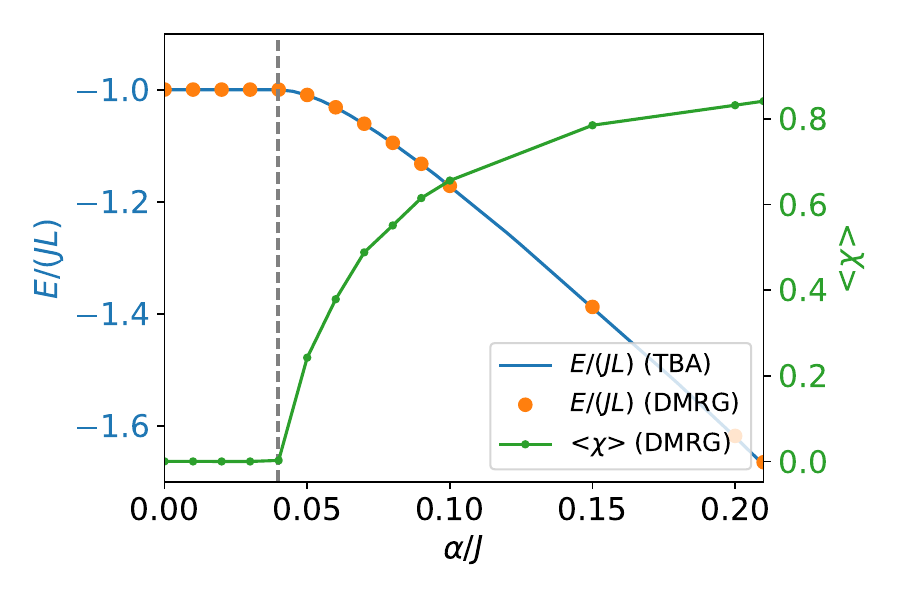}
  \put(35, 20){$\alpha_c/J = \frac{1}{8\pi}$}
  \put(34, 19){\vector(-1, -2){3}}
\end{overpic}
\caption{Ground-state energy density and chirality across the $Q_3$-driven transition. The blue curve is the exact TBA result for the dimensionless free energy, $f/J$, the orange points are DMRG energies for a periodic chain of length $L=100$, and the green points on the right axis show the magnitude of the DMRG chirality density. The energy is pinned at the undeformed BT value $f/J=-1$ up to $(\alpha_c/J)=1/(8\pi)$ (vertical dashed line) and then decreases smoothly. The chirality turns on at the same threshold, confirming that the postcritical sector is simultaneously current carrying and chiral.} 
\label{fig:energy-chi}
\end{figure}

Our main result is that the $Q_3$ tilt drives a sharp phase transition into a current-carrying chiral sector. Using the thermodynamic Bethe ansatz (TBA) and benchmarking the exact solution against DMRG, we show that the undeformed BT ground state remains unchanged up to
\begin{equation}
\alpha_c=\frac{J}{8\pi}
\;.
\end{equation}
For $\alpha < \alpha_c$, the tilt is present but too weak to reconstruct the rapidity sea, so the ground state remains the ordinary BT critical state with vanishing $\langle Q_3\rangle/N$ and vanishing $\chi$. For $\alpha>\alpha_c$, a finite rapidity interval forms, $\langle Q_3\rangle/N$ becomes nonzero, and the system enters a gapless chiral current-carrying sector that remains consistent with a $c=3/2$ conformal field theory. Near the threshold, the free energy starts quadratically in $\alpha-\alpha_c$, whereas the conserved current turns on linearly. DMRG shows that the scalar chirality turns on linearly at the same threshold, $\alpha_c$. The common onset of the energy response and the chirality is shown in Fig.~\ref{fig:energy-chi}. In this sense, the paper identifies--exactly--a spin-$1$ current-carrying critical sector that is selected by a conserved energy-current bias but diagnosed most transparently by a nonconserved chiral observable.

The paper is structured as follows. Sec.~\ref{sec:II} derives the conserved charges of the BT chain, makes the local structure of $Q_3$ explicit, and shows that the corresponding global charge is the physical energy current, up to the normalization fixed by Eq.~\eqref{eq:BT-H}. Sec.~\ref{sec:III} solves the tilted Hamiltonian $H_\alpha$ in the thermodynamic limit by TBA, establishes the exact threshold and the near-threshold equation of state, and benchmarks the analytical results against DMRG. Sec.~\ref{sec:IV} discusses experimental realizations, with an emphasis on optical-lattice platforms and programmable qutrit simulators. More broadly, the BT chain serves as a concrete demonstration of a general principle: a suitably chosen conserved quantity can convert a high-energy nonequilibrium problem into an equilibrium ground-state problem, while leaving ample freedom to probe the resulting sector with physically transparent observables.

\section{Conserved charges and energy current}
\label{sec:II}
Equations~\eqref{eq:hn-intro}, \eqref{eq:boost-intro}, and \eqref{eq:Q3-intro-def} already determine the algebraic origin of the third charge. Because the bond operator $h_n$ acts only on a single link, only adjacent bonds do not commute. Therefore, the first higher local charge is necessarily a translationally invariant three-site operator. In this section, we explicitly derive $Q_3$ and show that once the dimensional factor $J/4$ is restored, it is exactly the energy current of the spin-$1$ BT chain.

It is convenient to introduce the bilinear exchange operator on a bond,
\begin{equation}
A_n\equiv \mathbf{S}_n\cdot\mathbf{S}_{n+1}
\;,
\label{eq:An-def}
\end{equation}
so that $h_n=A_n-A_n^2$, and the local scalar chirality,
\begin{equation}
\chi_n\equiv
\mathbf{S}_n\cdot(\mathbf{S}_{n+1}\times \mathbf{S}_{n+2})
 \;,
\label{eq:chi-def}
\end{equation}
whose expectation value is the order parameter introduced in Eq.~\eqref{eq:chi-intro}. The key local commutator is
\begin{equation}
[A_n,A_{n+1}]=-i\chi_n
\;.
\label{eq:An-comm}
\end{equation}
Its physical meaning is transparent: the two bond exchanges share the spin at site $n+1$, and the noncommutativity of that shared spin records the handedness of the three-site configuration. Chirality therefore appears already at the first nontrivial step of the boost hierarchy.

The BT interaction is, however, bilinear-biquadratic rather than purely bilinear, so the full conserved charge is not exhausted by the bare scalar chirality. Expanding $h_n=A_n-A_n^2$ in Eq.~\eqref{eq:Q3-intro-def} gives
\begin{eqnarray}
&&[h_n,h_{n+1}]
=\\
&&[A_n,A_{n+1}]
-[A_n,A_{n+1}^2]
-[A_n^2,A_{n+1}]
+[A_n^2,A_{n+1}^2]
\;.\nonumber
\label{eq:hn-expand}
\end{eqnarray}
The remaining commutators are (see Appendix \ref{app:A} for details): 
\begin{align}
[A_n,A_{n+1}^2] &= -i\{\chi_n,A_{n+1}\}\;, \nonumber\\
[A_n^2,A_{n+1}] &= -i\{\chi_n,A_n\}\;, \nonumber\\
[A_n^2,A_{n+1}^2] &= -i\{A_n,\{A_{n+1},\chi_n\}\}
\;,
\label{eq:Q3-identities}
\end{align}
where $\{X,Y\}=XY+YX$ is the anticommutator. Substituting Eqs.~\eqref{eq:An-comm} and \eqref{eq:Q3-identities} into Eq.~\eqref{eq:Q3-intro-def}, we obtain
\begin{align}
Q_3 = -\sum_{n=1}^N \Bigl[
&\chi_n-\{\chi_n,(A_n+A_{n+1})\} \nonumber\\
&+\{A_n,\{A_{n+1},\chi_n\}\}
\Bigr]
\;.
\label{eq:Q3-explicit}
\end{align}
Thus, $Q_3$ is chiral, but not in the naive sense of being merely the sum of bare scalar chiralities. The first term is the symmetry-transparent chirality itself; the remaining terms arise from  the biquadratic part of the BT interaction and dress the current with additional local operators. It is convenient to collect these BT-specific corrections into
\begin{equation}
X_n\equiv
-\{\chi_n,(A_n+A_{n+1})\}
+\{A_n,\{A_{n+1},\chi_n\}\}
\;,
\label{eq:Xn-def}
\end{equation}
so that
\begin{equation}
Q_3=-\sum_{n=1}^N \left(\chi_n+X_n\right)
\;.
\label{eq:Q3-dressed}
\end{equation}
The new operator $X_n$ has the same transformation properties as $\chi_n$: it is invariant under global spin rotations and odd under both time reversal and spatial inversion. In this precise sense, $Q_3$ is a \emph{dressed chiral charge}. The conserved quantity selected by integrability is therefore a chiral current, but not simply the bare scalar chirality.


The same local structure also fixes the transport meaning of $Q_3$. Let
\begin{equation}
e_n=\frac{J}{4}h_n
\label{eq:bond-energy}
\end{equation}
denote the physical bond-energy density. Local energy conservation is written as
\begin{equation}
\dot e_n=i[H,e_n]=-(j_{n+1}^E-j_n^E)
 \;,
\label{eq:continuity}
\end{equation}
where $j_n$ are defined as the local current operators associated with the physical bond-energy density. Because only neighboring bond energies fail to commute, Eq.~\eqref{eq:hn-intro} gives
\begin{equation}
i[H,e_n]
=i\left(\frac{J}{4}\right)^2\Bigl([h_{n-1},h_n]-[h_n,h_{n+1}]\Bigr)
\;.
\label{eq:current-derivation}
\end{equation}
Comparing Eqs.~\eqref{eq:continuity} and \eqref{eq:current-derivation}, one finds
\begin{equation}
j_n^E=i\left(\frac{J}{4}\right)^2[h_{n-1},h_n]
\;.
\label{eq:local-current}
\end{equation}
(see Appendix~\ref{app:B} for details). Summing Eq.~\eqref{eq:local-current} over the chain and comparing with Eq.~\eqref{eq:Q3-intro-def}, we obtain
\begin{equation}
\mathcal{J}_E\equiv \sum_{n=1}^N j_n^E
=-\left(\frac{J}{4}\right)^2 Q_3
=-\frac{J}{4}\,i[\mathcal{B},H]
\;.
\label{eq:JE-Q3}
\end{equation}
With the minus-divergence convention of Eq.~\eqref{eq:continuity}, the overall sign in Eq.~\eqref{eq:JE-Q3} is fixed by the choice of current orientation. Reversing the orientation would flip both sides simultaneously but would not change the physical statement: up to the normalization inherited from Eq.~\eqref{eq:BT-H}, the third conserved charge is the energy current of the spin-$1$ BT chain.

The physical role of the tilt introduced in Eqs.~\eqref{eq:Halpha-intro} and \eqref{eq:q3-derivative-intro} is now transparent. The field $\alpha$ is conjugate to a conserved chiral current, so the deformation selects eigenstates with finite current and a preferred handedness. At the same time, Eqs.~\eqref{eq:Q3-explicit} and \eqref{eq:Q3-dressed} show why the scalar chirality in Eq.~\eqref{eq:chi-intro} is not a redundant observable. The tilt couples directly to the dressed current, not to the bare chirality alone. The fact that both $\langle Q_3\rangle/N$ and the scalar chirality turn on at the same critical value of $\alpha=\alpha_c$ is therefore a genuine dynamical result that is nontrivial.


\section{Current-carrying chiral eigenstates of the spin-$1$ Babujian-Takhtajan model}
\label{sec:III}
We now analyze the tilted Hamiltonian introduced in Eq.~\eqref{eq:Halpha-intro}. Because $[H,Q_3]=0$, varying $\alpha$ does not change the eigenbasis; it only reorders the energy levels. The ground state of $H_\alpha$ is therefore, in general, a highly excited eigenstate of the original BT Hamiltonian $H$, selected by the value of its conserved current quantum number. This is the central advantage of the construction: a nonequilibrium eigensector of $H$ is converted into an equilibrium ground-state problem for a local Hamiltonian.

The observable that diagnoses this selected sector most directly is the conserved charge density itself, obtained from the exact equation of state through Eq.~\eqref{eq:q3-derivative-intro}. The observable that reveals its internal handedness is the scalar chirality, already introduced in Eq.~\eqref{eq:chi-intro}. These two quantities play logically distinct roles: the tilt by $Q_3$ selects the sector, while the chirality tests whether the selected sector is genuinely chiral rather than merely current-carrying. This distinction is especially important in the spin-$1$ BT chain because, as shown in Sec.~\ref{sec:II}, $Q_3$ is a \emph{dressed} chiral current, not the bare scalar chirality alone.

\subsection{Thermodynamic Bethe ansatz for the tilted problem}

The Bethe ansatz provides an exact quasiparticle description of a one-dimensional integrable many-body system. The basic idea is that, because scattering factorizes into successive two-body processes, an eigenstate can be labeled by a set of rapidities $\{\lambda_j\}$, where each rapidity $\lambda_j$ is the spectral parameter that parametrizes the quasiparticle momentum $p_j$ as $\lambda_j=\cot{\frac{p_j}{2}}$, while the interactions are encoded exactly in the phase shifts that constrain the allowed rapidities. For the spin-$1$ BT chain these rapidities form the standard string patterns in the complex plane; in the thermodynamic limit, one trades individual roots for smooth densities of strings and holes. This is the point of the TBA: it provides the exact thermodynamics of the interacting rapidity quasiparticles. The explicit Bethe equations, the string construction, and the derivation of the thermodynamic equations are given in Appendix~\ref{app:C}; here we keep only the ingredients needed to discuss the results and the underlying physics. 

The first important observation is that the $Q_3$ deformation does not alter the integrable scattering structure. Since $Q_3$ commutes with $H$, the rapidity quantization conditions are the same as in the undeformed BT chain. What changes is the bare energy assigned to a given set of rapidities. For the tilted Hamiltonian, $H_{\alpha}$, in Eq.~\eqref{eq:Halpha-intro}, the single-particle contribution can be easily derived in the Bethe ansatz approach, yielding
\begin{equation}
E_\alpha=-\sum_{j=1}^{M}
\left(J-8\alpha\,\partial_{\lambda_j}\right)\frac{1}{1+\lambda_j^2}
\;.
\label{eq:sec3-energy}
\end{equation}
This formula makes the role of the tilt transparent. The first term is the usual BT dispersion. The second term, which is odd in rapidity, comes entirely from the conserved-current deformation and biases the occupation of the two rapidity tails in opposite ways. In real space, the field $\alpha$ couples to a conserved energy current; in rapidity space, it acts as an orientation-dependent driving term that favors one handedness of the Bethe sea over the other.

The next step is standard in the TBA. One passes from individual rapidities to particle and hole densities for the different string species and then minimizes the free energy at temperature $T$. The result is a set of self-consistent dressed energies $\varepsilon_n(\lambda)$, which measure the free-energy cost of adding an $n$-string with rapidity $\lambda$ to the equilibrium state. For the present model, the exact hierarchy is
\begin{align}
\varepsilon_1(\lambda)
&=
T\,s*\ln\!\left(1+e^{\varepsilon_2(\lambda)/T}\right)
\;,\nonumber\\
\varepsilon_n(\lambda)
&=
-\pi\left(J-8\alpha\,\partial_\lambda\right)s(\lambda)\,\delta_{n2}
\nonumber\\
&+T\,s*\ln\!\Big[
\left(1+e^{\varepsilon_{n-1}(\lambda)/T}\right)
\left(1+e^{\varepsilon_{n+1}(\lambda)/T}\right)
\Big],\nonumber\\
&n\ge 2
\;,
\label{eq:sec3-tba}
\end{align}
with $*$ denoting the convolution, and
\begin{equation}
s(\lambda)=\frac{1}{2\cosh(\pi\lambda)}
\;.
\label{eq:sec3-skernel}
\end{equation}
Eq.~\eqref{eq:sec3-tba} can be read as an exact self-consistent quasiparticle problem. The source term gives the bare energetic preference set by the microscopic Hamiltonian, while the convolution with $s(\lambda)$ describes how the surrounding sea dresses that bare cost through interactions. Negative dressed energy indicates that the corresponding rapidities tend to be occupied, while positive dressed energies are disfavored.

Two physical features of Eq.~\eqref{eq:sec3-tba} are particularly important. First, only the $n=2$ sector is driven directly. This is the thermodynamic fingerprint of the antiferromagnetic spin-$1$ BT chain: its ground-state sea is built from two-strings, while higher strings contribute only through dressing. Second, the $Q_3$ deformation appears only in that bare two-string source. Consequently, the tilt does not modify the scattering kernel or the allowed string content; it simply selects which part of the same integrable spectrum becomes the ground state of $H_\alpha$. In this sense, the transition discussed below corresponds to a reorganization of the occupied rapidity sea within the TBA and as such, is of Lifshitz universality class.

This perspective also clarifies the zero-temperature limit. As $T\to0$, the thermodynamics reduces to a question of signs: the equilibrium state is determined by which portions of the dressed two-string dispersion lie below or above zero. In this limit, the entire problem therefore collapses to the sign structure of a single dressed-energy function, which can then be analyzed exactly. We now turn to that zero-temperature equation of state. Technical details of the string hypothesis, the density equations, and the derivation of Eq.~\eqref{eq:sec3-tba} are given in Appendix~\ref{app:C}.

\subsection{Zero-temperature equation of state and the onset of current}
\begin{figure}
    \centering  \includegraphics[width=1\linewidth]{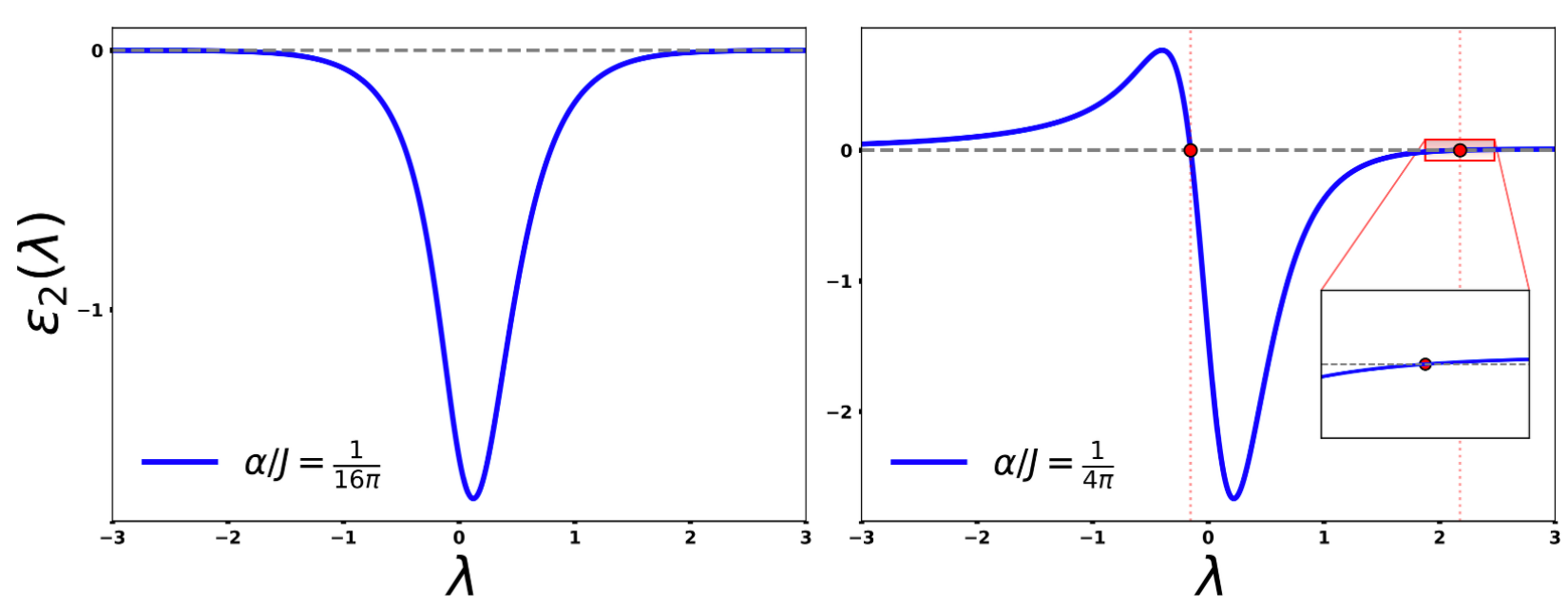}
    \caption{
    Dressed two-string energy $\varepsilon_2(\lambda)$ below and above the current-driven transition. Left panel: $\alpha/J=1/(16\pi)<\alpha_c/J$, for which $\varepsilon_2(\lambda)<0$ across the entire real axis and the BT phase has no Fermi boundaries. Right panel: $\alpha/J=1/(4\pi)>\alpha_c/J$, for which $\varepsilon_2(\lambda)$ has two simple zeros at $b_-$ and $b_+$ (vertical dashed lines and red dots), with $\varepsilon_2(\lambda)<0$ only on the interval $b_-<\lambda<b_+$. The steep left crossing and the much shallower right crossing imply that the two Fermi edges of the current-carrying chiral phase are inequivalent. The inset zooms in on the neighborhood of $b_+$ and confirms that the right zero remains simple.
        }
    \label{fig:two-boundary}
\end{figure}

The state of interest in this section is the ground state of $H_\alpha$ at $T\to0$. In this limit, the TBA reduces to a single nonlinear integral equation for the dressed energy of the two-string sector,
\begin{equation}
\varepsilon_2(\lambda)=g(\lambda)+
\int_{-\infty}^{\infty}
R(\lambda-\mu)\,\varepsilon_2^+(\mu)\,d\mu
\;,
\label{eq:sec3-eps2full}
\end{equation}
with
\begin{align}
   & g(\lambda)=
-\pi\left(J-8\alpha\,\partial_\lambda\right)s(\lambda)
\;, \nn
\\
&R(\lambda)= (s*s+s*a_1)(\lambda)
\;,
\label{eq:sec3-gR}
\end{align}
and
\begin{equation}
a_n(\lambda)=
\frac{1}{2\pi}
\frac{n}{\lambda^2+\left(\frac{n}{2}\right)^2}
\;.
\label{eq:sec3-an}
\end{equation}
At zero temperature the sign of $\varepsilon_2(\lambda)$ fully determines the selected macrostate: regions where $\varepsilon_2(\lambda)<0$ are occupied, while regions with $\varepsilon_2(\lambda)>0$ are empty. The problem therefore reduces entirely to the sign structure of one dressed-energy function.

That sign structure is controlled by the explicit source (driving) term:
\begin{equation}
g(\lambda)=
-\frac{\pi J}{2}\operatorname{sech}(\pi\lambda)
\left[
1+8\pi\frac{\alpha}{J}\tanh(\pi\lambda)
\right]
\;.
\label{eq:sec3-gexplicit}
\end{equation}
The even part of the source would fill a symmetric two-string sea. The odd part, produced solely by the conserved tilt, destabilizes the sea first in one rapidity tail. This is the exact thermodynamic signature of a current-carrying state: the deformation leaves the center of the distribution unaffected, creating instead an occupation imbalance between the two tails.

The threshold is then obtained by identifying when the bracket in Eq.~\eqref{eq:sec3-gexplicit} first changes sign, which occurs at $\alpha=\alpha_c$.
For $\alpha\le \alpha_c$, the kernel in Eq.~\eqref{eq:sec3-eps2full} cannot generate a positive part of the dressed energy from a nonpositive source. The selected state is therefore still the undeformed BT ground state, and the exact equation of state remains flat,
\begin{equation}
f=-J
\;,
\qquad
\frac{1}{N}\langle Q_3\rangle=0
\;,
\qquad
\alpha\le \alpha_c
\;.
\label{eq:sec3-flat-phase}
\end{equation}
This regime is nevertheless nontrivial: although the deformation is present, it remains too weak to reorder the spectrum. In particular, no current-carrying eigenstate has yet been pulled below the original BT ground state.

For $\alpha>\alpha_c$ the structure changes qualitatively. The dressed energy develops two zeros,
\begin{align}
&\varepsilon_2(\lambda)<0
\quad \text{for } \lambda\in [b_-,b_+]\;,\nonumber\\
&\varepsilon_2(\lambda)>0
\quad \text{for } \lambda\notin [b_-,b_+] 
\;,
\label{eq:sec3-sign}
\end{align}
so that the occupied two-string sea no longer extends over the entire rapidity axis but instead becomes confined to a single finite interval. Because the source is asymmetric, this interval is shifted away from the origin. The postcritical macrostate is therefore a current-carrying Fermi sea with two self-consistently determined edges.

Figure~\ref{fig:two-boundary} provides the most straightforward rapidity-space view of this reconstruction. For $\alpha/J=1/(16\pi)$, which is below the threshold, $\varepsilon_2(\lambda)$ remains negative for all $\lambda$, so the two-string sea occupies the entire rapidity axis and there are no Fermi boundaries. For $\alpha/J=1/(4\pi)$, which lies inside the current-carrying chiral phase, $\varepsilon_2(\lambda)$ turns positive in both asymptotic tails and vanishes at two distinct rapidities $b_-<b_+$. The occupied rapidities are therefore restricted to a single connected interval $b_-<\lambda<b_+$, which is the sharp thermodynamic signature of the postcritical phase.

The key feature of the present solution is not just the presence of two Fermi points but the fact that they are strongly inequivalent. In many standard integrable TBA ground states, symmetry imposes a parity-even interval $[-B, B]$, with the left and right boundaries being mirror images. Here, however, the odd driving term in Eq.~\eqref{eq:sec3-gexplicit} breaks the $\lambda\to-\lambda$ symmetry already at the bare level. The left boundary $b_-$ is the main instability: it appears when the negative-rapidity tail first changes sign at $\alpha_c$. The right boundary $b_+$ is then generated self-consistently by the dressing kernel, ensuring the occupied sea stays connected. Thus, the chiral phase is characterized not by a symmetric two-Fermi-point sea or a one-edge Wiener-Hopf problem but by an inherently asymmetric two-boundary state.

In the postcritical regime the dressed-energy problem becomes genuinely two-boundary, with the occupied sea bounded by the Fermi points $b_-$ and $b_+$. At zero-temperature, the dressed energy on the occupied interval satisfies the exact equation
\begin{eqnarray}
&&\varepsilon_2(\lambda)
+\int_{b_-}^{b_+}
K(\lambda-\mu)\,\varepsilon_2(\mu)\,d\mu
=
-\pi\left(J-8\alpha\,\partial_\lambda\right)q(\lambda)\;,\nonumber\\
&&\lambda\in[b_-,b_+]
\;,
\label{eq:sec3-two-boundary}
\end{eqnarray}
where
\begin{equation}
q(\lambda)=a_1(\lambda)+a_3(\lambda)\;,
\quad
K(\lambda)=2a_2(\lambda)+a_4(\lambda)\;,
\label{eq:sec3-qK}
\end{equation}
and
\begin{equation}
\varepsilon_2(b_-)=\varepsilon_2(b_+)=0
\;,
\label{eq:sec3-boundaries}
\end{equation}
where the interval $[b_-,b_+]$ emerges from the full dressed-energy problem. Therefore, the ground-state energy density follows from
\begin{equation}
f=
\int_{b_-}^{b_+} q(\lambda)\,\varepsilon_2(\lambda)\,d\lambda
\;.
\label{eq:sec3-free}
\end{equation}
Together with Eq.~\eqref{eq:q3-derivative-intro}, this relation determines the exact equation of state of the current-resolved sectors. Close to the threshold, one finds a one-sided onset, 
\begin{align}
&\frac{f}{J}
=
-1
-\frac{32\pi^4}{27}
\left(\frac{\alpha-\alpha_c}{J}\right)^2
+O\!\left[\left(\frac{\alpha-\alpha_c}{J}\right)^3\right]\;,
\nonumber\\
&\frac{1}{N}\langle Q_3\rangle
=
-\frac{64\pi^4}{27}
\left(\frac{\alpha-\alpha_c}{J}\right)
+O\!\left[\left(\frac{\alpha-\alpha_c}{J}\right)^2\right]
\;.
\label{eq:sec3-onset}
\end{align}
The free energy therefore starts quadratically above $\alpha_c$, while the conserved current turns on linearly. This is the precise thermodynamic content of the tail-instability picture: a Fermi boundary enters from $-\infty$, so the phase-space weight of the emptied tail is linear in $\alpha-\alpha_c$, and the current responds at the same order.

The unequal tangent angles seen in Fig.~\ref{fig:two-boundary} reveal additional physics. Near $b_-$, the dressed energy crosses zero quite abruptly, while near $b_+$ the crossing is much more gradual. In rapidity space, a small displacement away from $b_-$ changes the excitation energy more strongly than the same displacement away from $b_+$. The real low-energy excitations in the chiral phase are particle-hole excitations created near these two Fermi rapidities. Because the two edges are not related by parity, the corresponding quasiparticle branches are inherently different. Strictly speaking, the physical propagation velocities are obtained only after converting rapidity to momentum through the usual Bethe-ansatz Jacobian, so the slopes in Fig.~\ref{fig:two-boundary} should not be directly identified with the sound velocities. Even so, the different tangent angles already indicate varying local energy scales and different low-energy phase-space weights for excitations created near $b_-$ and $b_+$. In this sense, the finite current is carried by an unbalanced pair of gapless Fermi-edge excitations rather than by two symmetry-related modes.

\subsection{DMRG benchmarks, chirality, and criticality}

The TBA describes the thermodynamic limit. To demonstrate that the same TBA structure is already visible directly at the microscopic level for finite-size systems, we compared the analytical TBA predictions with the DMRG results for a chain of length $L=100$ with a periodic boundary condition.
 This comparison is particularly natural because the tilted Hamiltonian remains local and one-dimensional. Standard ground-state DMRG therefore targets the desired state directly, without requiring access to the excited states of the original Hamiltonian. 

\begin{figure}[t]
    \centering
    \begin{overpic}[width=\linewidth]{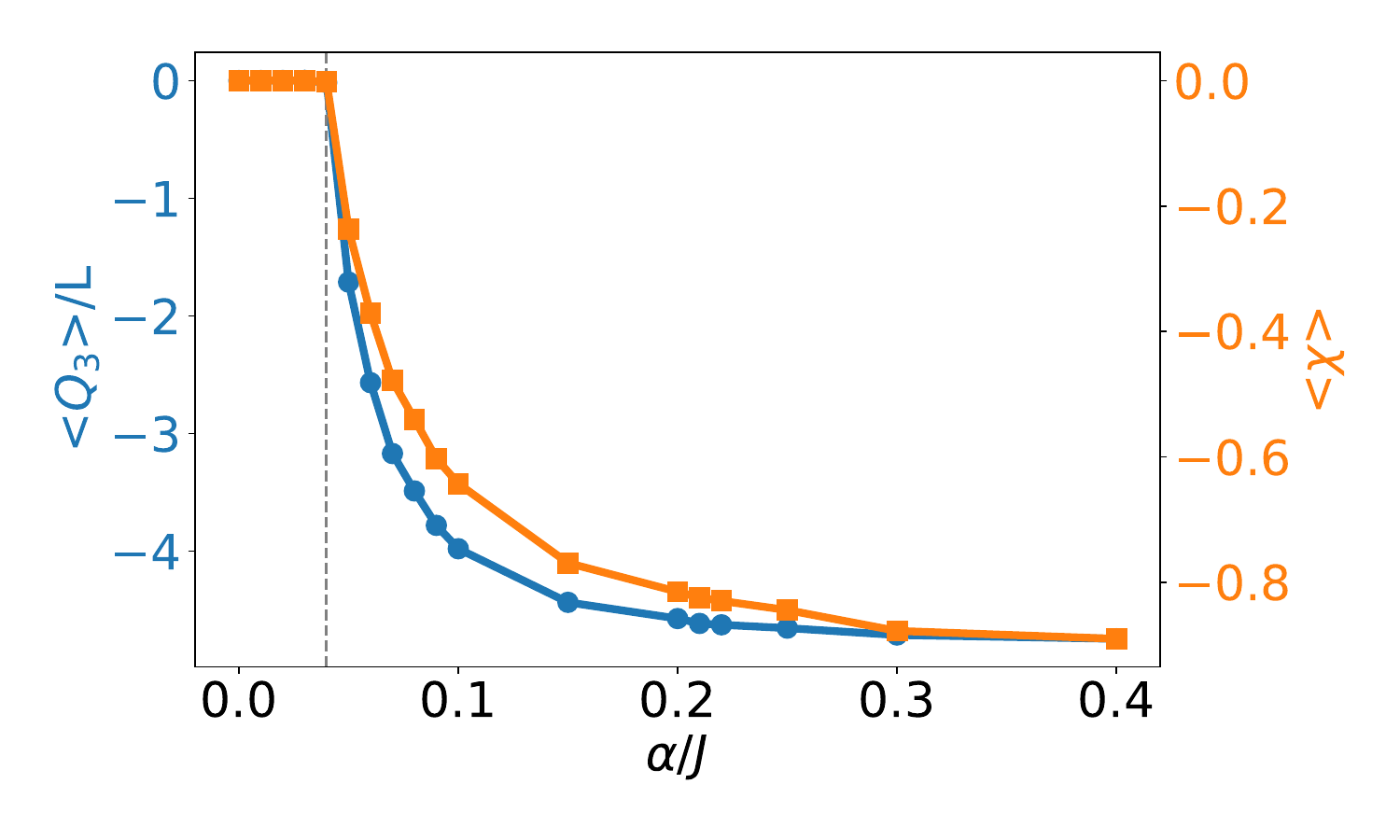}
      \put(34, 32){$\alpha_c/J = \frac{1}{8\pi}$}
      \put(33, 31){\vector(-1, -2){9}}
    \end{overpic}
    \caption{Conserved-current density and chirality density as functions of $\alpha/J$. Both quantities vanish in the BT phase and become finite at the same critical value, $\alpha_c/J=1/(8\pi)$, shown by the dashed line.}
    \label{fig:q3-chi}
\end{figure}


The comparison between the exact two-boundary TBA energy density and the DMRG ground-state energy, on one hand, and the magnitude of the DMRG chirality density, on the other, is shown in Fig.~\ref{fig:energy-chi}. The agreement of the energy curve is excellent on both sides of the transition. Two points are particularly  important. First, the energy remains pinned at $f/J=-1$ up to the exact threshold, $\alpha_c$, confirming the flat BT phase, Eq.~\eqref{eq:sec3-flat-phase}. Second, immediately above the threshold the DMRG chirality turns on at the same point where the energy departs from its BT value. This common onset provides the first direct evidence that the postcritical sector is not only current-carrying but also genuinely chiral.

The onset at $\alpha_c$ is more explicit in Fig.~\ref{fig:q3-chi}, which shows both the conserved-current density and the chirality density. With the sign convention fixed in Sec.~\ref{sec:II}, positive $\alpha$ selects negative $\langle Q_3\rangle/N$, and the chirality follows the same orientation. The rapid initial growth is consistent with the linear threshold law in Eq.~\eqref{eq:sec3-onset}, while the broader crossover at larger $\alpha/J$ reflects the full nonlinear rearrangement of the occupied interval $[b_-,b_+]$.

\begin{figure}[t]
    \centering
    \includegraphics[width=\linewidth]{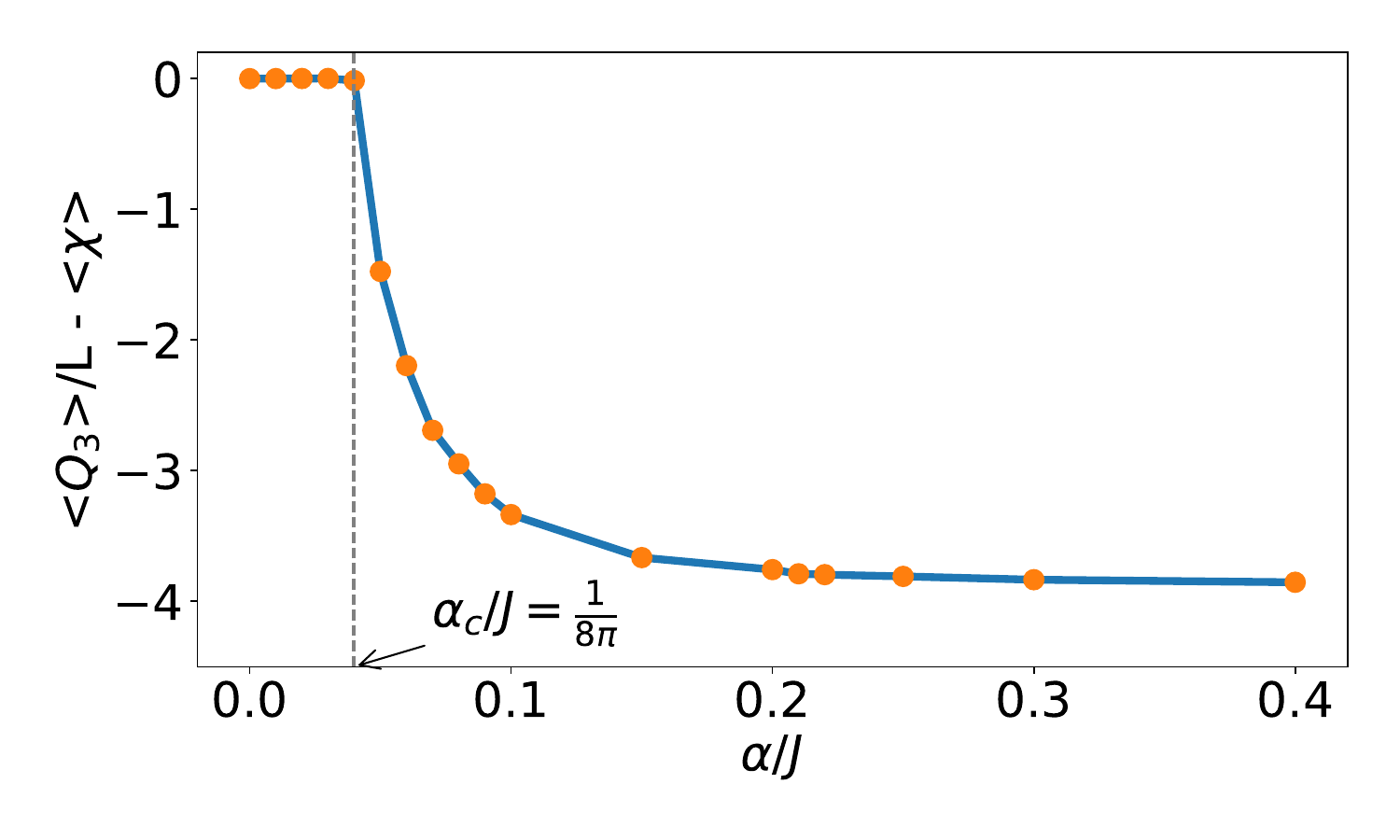}
    \caption{Difference between the conserved-current density and the chirality density. The difference remains zero below $\alpha_c$ because both quantities vanish in the flat BT phase, but becomes finite immediately above the transition. This directly demonstrates that $Q_3$ is not the bare scalar chirality alone: the BT-specific dressing terms identified in Sec.~\ref{sec:II} contribute throughout the current-carrying sector.}
    \label{fig:q3-minus-chi}
\end{figure}

Figure~\ref{fig:q3-minus-chi} addresses a different question: whether the conserved current is merely a bare scalar chirality in disguise. It is not. The difference between the two densities remains zero throughout the flat BT phase. This is because both quantities vanish in this region, but it becomes finite immediately above $\alpha_c$ and stays finite throughout the current-carrying regime. This behavior is in precise agreement with the predictions of Sec.~\ref{sec:II}. The extra BT-specific dressing terms in the local density of $Q_3$ are therefore not negligible corrections; they make an $O(1)$ contribution in the postcritical sector. 


\begin{figure}[t]
    \centering
    \includegraphics[width=\linewidth]{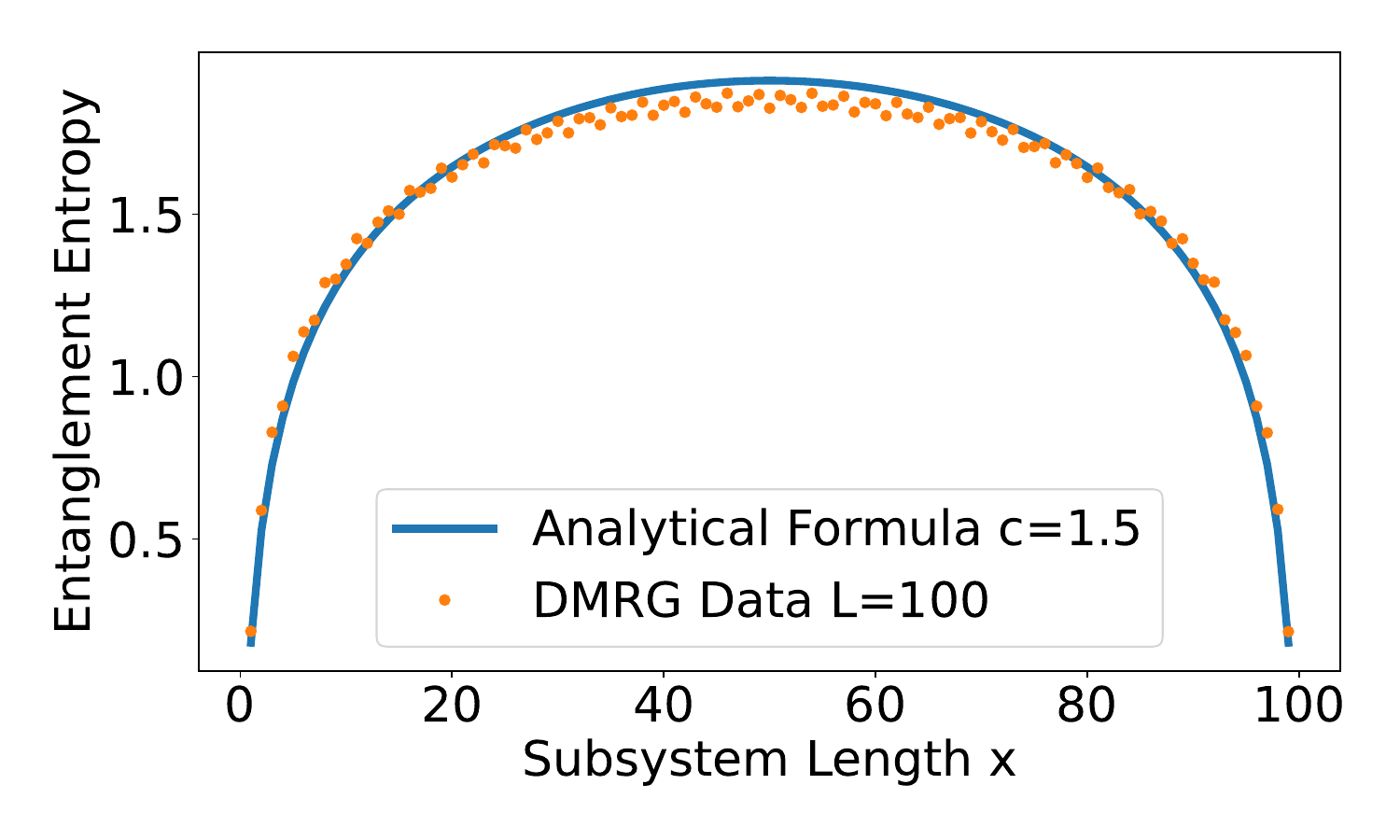}
    \caption{Entanglement entropy of the $Q_3$-deformed chain at $\alpha/J=0.1$ for a periodic system of length $L=100$. The DMRG data follow the standard conformal profile with central charge $c=3/2$. The postcritical current-carrying sector is therefore still gapless and consistent with a $c=3/2$ conformal field theory.}
    \label{fig:entropy}
\end{figure}

A second key question is whether the postcritical sector remains critical. The answer is yes.  At $\alpha/J=0.1$, well inside the current-carrying regime, the DMRG simulations in Fig.~\ref{fig:entropy} show the entanglement entropy following the standard open-chain conformal profile~\cite{CalabreseCardy} with central charge $c=3/2$. Thus, the $Q_3$ tilt reorganizes the occupied rapidity interval and generates a finite current and chirality without opening a gap. The natural interpretation is that the deformation selects a different \emph{gapless} eigensector of the BT chain, consistent with the $c=3/2$ conformal description.

Taken together, the five figures establish the main results of the paper. Fig.~\ref{fig:energy-chi} confirms the exact threshold and the flat phase. Fig.~\ref{fig:two-boundary} provides the rapidity-space view of the transition and makes the asymmetric two-boundary structure explicit. Fig.~\ref{fig:q3-chi} shows that the conserved current and the chirality turn on together and that the response extends smoothly into the postcritical regime. Fig.~\ref{fig:q3-minus-chi} demonstrates that the conserved current is a dressed, rather than a bare, chiral observable. Fig.~\ref{fig:entropy} indicates that the postcritical sector remains critical with a central charge $c=3/2$. Collectively, these results present a clear picture: for $\alpha\le\alpha_c$, the selected state matches the ordinary BT ground state, whereas for $\alpha>\alpha_c$, the system enters a current-carrying, chiral, and still gapless eigensector of the original spin-$1$ BT chain.


\section{Experimental Platforms}
\label{sec:IV}

A practical advantage of the present construction is that the experiment does not need to maintain an open-system energy flow. The target is the static local Hamiltonian
\begin{equation}
H_\alpha = H + \alpha Q_3
= \frac{J}{4}\sum_n h_n + \alpha \sum_n q\strut_{3,n}
\;,
\end{equation}
where $q\strut_{3,n}$ is the local three-site density derived in Sec.~\ref{sec:II}. Because $[H,Q_3]=0$, every eigenstate of $H_\alpha$ is also an eigenstate of $H$. In particular, if $|\psi_0(\alpha)\rangle$ is the exact ground state of $H_\alpha$, then turning off the bias does not produce intrinsic dynamics under the original BT Hamiltonian:
\begin{equation}
e^{-itH}\,|\psi_0(\alpha)\rangle = e^{-itE}\,|\psi_0(\alpha)\rangle
\;.
\end{equation}
This is the operational distinction between the present proposal and a conventional transport experiment: the current-carrying sector is prepared as a ground state and then diagnosed by static measurements, with the optional release protocol providing an additional check of whether the implemented tilt is really conserved.

The two experimentally most relevant settings are complementary. Programmable qutrit (a 3-level quantum system described by a superposition of three orthogonal states) simulators are the most natural route to the exact local operator content of $H+\alpha Q_3$ on finite chains. Optical lattices are the most natural route to long one-dimensional spin-$1$ systems, but in that setting, neither the BT point nor the full-dressed current $Q_3$ emerges automatically from standard superexchange.

\subsection{Programmable qutrit simulators}

The local Hilbert space of a spin-$1$ chain is exactly a qutrit, so the BT model is native to hardware that manipulates three-level systems. On the trapped-ion side, integer-spin chains were realized in Ref.~\cite{Senko2015}, native qudit entangling gates were demonstrated in Ref.~\cite{Hrmo2023}, and three- and four-body interactions were demonstrated in Ref.~\cite{Katz2023}. Closely related qutrit many-body physics was realized in Ref.~\cite{Edmunds2025}. On the superconducting side, high-fidelity two-qutrit entangling gates were reported in Refs.~\cite{Goss2022,LuoQutrit2023}, while multibody interactions relevant to chiral dynamics and operator synthesis were demonstrated in Refs.~\cite{Liu2020,Zhang2022}. These works, by themselves, do not realize the Hamiltonian studied here, but they establish the basic experimental ingredients needed for finite-size implementations.

For the present problem, the important structural point is locality. The target Hamiltonian is
\begin{equation}
H_\alpha
=
\frac{J}{4}\sum_n h_n
+
\alpha \sum_n q\strut_{3,n}
\;,
\end{equation}
with $h_n$ acting on two neighboring qutrits and $q^{}_{3,n}$ acting on three neighboring qutrits. This immediately implies that the conserved tilt does not create a conceptual obstruction for digital compilation. At the level of the exact target Hamiltonian,
\begin{equation}
e^{-i\delta t H_\alpha}=e^{-i\delta t H}\,e^{-i\delta t\alpha Q_3}
\;,
\end{equation}
exactly, because $[H,Q_3]=0$. The only approximation enters when the two commuting global factors are themselves decomposed into local gates. For example,
\begin{align}
e^{-i\delta t H}
&=
\prod_n e^{-i\delta t \frac{J}{4} h_n}
+O(\delta t^2)
\;,
\nonumber\\
e^{-i\delta t\alpha Q_3}
&=
\prod_n e^{-i\delta t\alpha q^{}_{3,n}}
+O(\delta t^2)
\;.
\end{align}
Thus, in a gate-based implementation, deviations from exact conservation arise as ordinary Hamiltonian-synthesis errors from the local compilation, not from any noncommutativity between $H$ and $Q_3$ themselves. This is the way to interpret finite-step effects in the present problem.

The most direct observables are the dressed current density and the scalar chirality. The operator definition
\begin{equation}
\frac{\langle Q_3\rangle}{N}
=
\frac{1}{N}\sum_n \langle q^{}_{3,n}\rangle
\;,
\end{equation}
is independent of how the state was prepared. For the exact target Hamiltonian, one also has the Hellmann-Feynman identity
\begin{equation}
\frac{\langle Q_3\rangle}{N}
=
\frac{1}{N}\frac{\partial E_0(\alpha)}{\partial \alpha}
\;,
\end{equation}
where $E_0(\alpha)$ is the ground-state energy of $H_\alpha$. In an analog realization, this relation is exact. In a digital realization it is recovered in the limit that the compiled dynamics converges to the target Hamiltonian, whereas the direct expectation value of $q_{3,n}$ remains the cleaner operational observable at finite step size. The independently measured chirality density
\begin{equation}
\chi
=
\frac{1}{N}\sum_n \langle \chi_n\rangle
\;,
\end{equation}
then tests the central physical statement of the paper: the postcritical sector is chiral, but the conserved tilt couples to a dressed current rather than to the bare scalar chirality alone.

For a finite chain, the two signatures remain sharp. The first is the common onset of $\langle Q_3\rangle/N$ and $\chi$ as $\alpha$ is increased. The second is the release test. If the prepared state is a high-fidelity eigenstate of the original Hamiltonian $H$, then switching off the bias should leave the state stationary up to experimental imperfections. Any residual post-release dynamics then quantifies the deviation of the implemented Hamiltonian from the ideal conserved tilt. In this sense, the release protocol is best viewed as a stringent fidelity benchmark for the exact operator synthesis.

The locality of $h_n$ and $q_{3,n}$ makes programmable trapped-ion and superconducting platforms the clearest route to finite-system realizations of the exact Hamiltonian studied in this work. What is experimentally demanding is not the size of the local Hilbert space, but the accurate synthesis of the three-site dressed current density.

\subsection{Optical lattices}

Optical lattices provide the complementary analog route to long spin-$1$ chains. Two closely related implementations are relevant. One uses genuine spin-$1$ bosons in an $F=1$ hyperfine manifold. The other uses composite bosons or two-particle local triplet sectors as effective qutrits. The general superexchange framework for spin models in optical lattices was established in Refs.~\cite{Duan2003,GarciaRipoll2004}. The spin-$1$ bilinear-biquadratic problem for spinor bosons was analyzed in Ref.~\cite{Imambekov2003}. Tunable single-ion anisotropy in a spin-$1$ setting was demonstrated in Ref.~\cite{Chung2021};, superexchange dynamics of composite spin-$1$ bosons was studied in Ref.~\cite{LuoSpin12024}, and Haldane physics in a related cold-atom setting was realized in Ref.~\cite{Sompet2022}. These results establish that long one-dimensional spin-$1$ systems and local spin diagnostics are experimentally realistic in cold-atom platforms.

A convenient starting point is the spinor Bose-Hubbard model
\begin{eqnarray}
H_{\rm BH}
&=&
-t \sum_{\langle ij\rangle,m}
\left(
 b^\dagger_{im} b_{jm} + \text{H.c.}
\right)
+
\frac{U}{2}\sum_i n_i(n_i-1)\nonumber\\
&+&
\frac{U_s}{2}\sum_i \left(\mathbf{S}_i^2-2n_i\right)
+
D\sum_i (S_i^z)^2
\;,
\end{eqnarray}
which, in the Mott regime $t\ll U,|U_s|$, generates an effective spin model through virtual doublon processes. Using the standard second-order strong-coupling expansion, one obtains
\begin{equation}
U_{F=0}=U-2U_s
\;,
\qquad
U_{F=2}=U+U_s
\;,
\end{equation}
\begin{equation}
H_{\rm eff}^{(2)}
=
-4t^2 \sum_{\langle ij\rangle}
\left(
\frac{P^{(0)}_{ij}}{U_{F=0}}
+
\frac{P^{(2)}_{ij}}{U_{F=2}}
\right)
+
D\sum_i (S_i^z)^2
\;,
\end{equation}
which can be rewritten as the familiar bilinear-biquadratic Hamiltonian
\begin{equation}
H_{\rm eff}^{(2)}
=
\sum_{\langle ij\rangle}
\left[
J_1\, \mathbf S_i\!\cdot\!\mathbf S_j
+
J_2\, (\mathbf S_i\!\cdot\!\mathbf S_j)^2
\right]
+
D\sum_i (S_i^z)^2
+
\text{const}
\;,
\end{equation}
\begin{equation}
J_1=-\frac{2t^2}{U_{F=2}}
\;,
\qquad
J_2=-\frac{4t^2}{3U_{F=0}}-\frac{2t^2}{3U_{F=2}}
\;.
\end{equation}
This standard derivation already shows the main limitation of the static optical-lattice route: the generic Mott insulator produces a broad bilinear-biquadratic family, not the integrable BT point automatically. The exact ratio of couplings required by the BT chain, therefore, has to be engineered rather than assumed.

The same caution is even more important for the current bias. In loop geometries with artificial gauge flux, third-order virtual processes can generate three-site chiral interactions. The spin-$1/2$ case reduces to a scalar-chirality term of the type discussed in Ref.~\cite{Pachos2004}. For spin-$1$ systems, however, one should \emph{not} identify the third-order effective Hamiltonian with that single operator. Additional three-site terms generally appear at the same order, and their coefficients depend on the microscopic realization. For this reason, we do not claim that a generic flux-assisted spin-$1$ optical lattice realizes the exact deformation $H+\alpha Q_3$ studied in this paper. Establishing such a realization would require a dedicated microscopic derivation for the specific driven or flux-threaded setup.

However, optical lattices are well suited to probing long-chain spin-$1$ physics, local correlators, and chirality-related observables. The scalar chirality can be accessed through three-site correlation measurements after appropriate local spin manipulations. Reconstructing the full dressed current density $q_{3,n}$ is more demanding, because it requires the additional bilinear-biquadratic dressing terms derived in Sec.~\ref{sec:II}. Thus, in cold atoms, the most realistic near-term goal is to explore related chiral spin-$1$ physics and finite-size or finite-temperature trends, rather than an exact realization of the integrable conserved tilt.

The roles of the two platforms are therefore complementary. Programmable qutrit devices are the cleanest route to the exact local Hamiltonian $H+\alpha Q_3$ on finite systems and to the stationary release test. Optical lattices are the natural route to longer one-dimensional chains and to analog studies of related chiral spin-$1$ physics, but the exact BT couplings and the full dressed current bias remain nontrivial engineering targets. 

\section{Conclusion}
\label{sec:V}
In this work, we used the spin-$1$ BT chain to show that a conserved-charge deformation can convert a high-energy nonequilibrium sector of an interacting model into the ground state of a local Hamiltonian. The central operator is the third conserved charge of the boost hierarchy, Eq.~\eqref{eq:Q3-intro-def}. We established two key facts: its local density is a dressed chiral operator rather than the bare scalar chirality alone, and its global form is the physical energy current (up to the normalization). The tilted Hamiltonian in Eq.~\eqref{eq:Halpha-intro} therefore provides a local, and exactly controlled current bias that leaves the eigenstates of $H$ unchanged while reordering them according to their current quantum number. This is the basic conceptual result of the paper.

We then solved the tilted problem in the thermodynamic limit using the TBA and benchmarked the solution against DMRG. The exact equation of state shows that the undeformed BT ground state survives up to the threshold, $\alpha_c$. Equivalently, the flat phase in Eq.~\eqref{eq:sec3-flat-phase} persists even though the current bias is already present. For $\alpha>\alpha_c$, the dressed two-string sea reconstructs into the finite interval $[b_-,b_+]$, and the system enters a current-carrying sector with finite $\langle Q_3\rangle/N$. The exact near-threshold behavior shows that the free energy starts quadratically, whereas the conserved current and the chirality turn on linearly. Physically, the instability begins in the far-left rapidity tail: a Fermi boundary enters from $-\infty$, and the resulting occupation imbalance produces a linear onset of the current.

The numerical results sharpen this picture and confirm that the postcritical sector is genuinely chiral. We see the common onset of the energy response, the conserved current, and the scalar chirality at the same $\alpha_c$. We demonstrated that $\langle Q_3\rangle/N$ and $\chi$ are not identical observables in the spin-$1$ chain, exactly because the local density of $Q_3$ is dressed by the biquadratic interaction. We also showed that the entanglement entropy in the postcritical regime remains consistent with a central charge $c=3/2$. The phase selected by the $Q_3$ tilt is therefore a gapless, current-carrying chiral sector of the original BT Hamiltonian.

These results open several concrete directions. One direction is to extend the analysis from $Q_3$ to higher conserved charges and determine which additional nonequilibrium eigensectors can be accessed by analogous local conserved tilts. Another direction is to develop the field-theory description of the postcritical sector more explicitly, including its correlation functions, operator content, and response to boundary conditions and finite-size quantization. A third direction is to study the fate of the chiral current-carrying phase under integrability-breaking perturbations, where one can ask which features survive as prethermal or long-lived signatures once the exact conservation law is relaxed. A fourth direction is to connect the present construction to nonequilibrium protocols, such as quenches, generalized Gibbs ensembles, and generalized hydrodynamics~\cite{GHD1,GHD2,Doyon-2025}, where the exact current bias provides a natural way to prepare states with controlled conserved-charge densities. Finally, the distinction between dressed current and bare chirality should be especially important for experiments. As discussed in Sec.~\ref{sec:IV}, programmable qutrit architectures appear well suited for implementing the exact local density of $Q_3$, whereas optical lattices are promising for probing the same physics in longer chains. In both settings, the simultaneous onset of $\langle Q_3\rangle/N$ and $\chi$, together with the stationary release protocol for the exact tilt, provides a clear target for future measurements.

More broadly, the paper illustrates a general principle: exact conserved quantities can be used not only to classify integrable spectra, but also to make otherwise inaccessible high-energy quantum phases amenable to ground-state methods. In the spin-$1$ BT chain, this principle exposes a chiral current-carrying critical sector with an exact threshold and an exact near-threshold equation of state. We expect the same strategy broadly applicable in systems where nontrivial conserved quantities are known explicitly.


\section*{Acknowledgments}
The authors thank Junjun Pang and Parameshwar R. Pasnoori for valuable discussions. The authors gratefully acknowledge support from the Simons Center for Geometry and Physics, Stony Brook University, where part of the research for this paper was performed. T.~A.~Sedrakyan also gratefully acknowledges support from the Max-Planck-Institut f{\"u}r Physik komplexer Systeme, where part of the research for this paper was performed.
The research was supported by the Armenian Higher Education and Science Committee under the ARPI Remote Laboratory program 24RL-1C024, research projects 21AG-1C024 and 25Post- Doc1C003.
\appendix
\renewcommand\thesubsection{\thesection.\arabic{subsection}}

\section{Derivation of the local form of \texorpdfstring{$Q_3$}{TEXT}}
\label{app:A}

Starting from Eq.~\eqref{eq:An-def}, the basic commutator is obtained directly from the spin algebra on the common site:
\begin{align}
[A_n,A_{n+1}]
&=
\sum_a S_n^a\,[S_{n+1}^a,\mathbf{S}_{n+1}\!\cdot\!\mathbf{S}_{n+2}]\nonumber\\
&=
-i\sum_a S_n^a\,(\mathbf{S}_{n+1}\times\mathbf{S}_{n+2})^a
\;.
\end{align}
The last line is precisely the scalar chirality defined in Eq.~\eqref{eq:chi-def}, so it reproduces Eq.~\eqref{eq:An-comm}. This is the microscopic origin of chirality in the third charge: two neighboring exchange terms try to rotate the middle spin in incompatible ways, and their commutator remembers the orientation of the three-spin triad.

 Using the product rule for commutators together with Eq.~\eqref{eq:An-comm}, one finds
\begin{align}
[A_n,A_{n+1}^2]
&=
[A_n,A_{n+1}]A_{n+1}+A_{n+1}[A_n,A_{n+1}],
\nonumber\\
[A_n^2,A_{n+1}]
&=
A_n[A_n,A_{n+1}]+[A_n,A_{n+1}]A_n,
\nonumber\\
[A_n^2,A_{n+1}^2]
&=
A_n[A_n,A_{n+1}^2]+[A_n,A_{n+1}^2]A_n.
\end{align}
Substituting Eq.~\eqref{eq:An-comm} into these three lines gives Eq.~\eqref{eq:Q3-identities} immediately.

\section{\texorpdfstring{$Q_3$}{TEXT} as the energy current of the spin-\texorpdfstring{$1$}{TEXT}  BT chain}

\label{app:B}

 A lattice current is fixed by the continuity equation, so it is controlled by the commutator of neighboring bond energies. Since each bond energy acts on two sites, the current necessarily acts on three consecutive sites, exactly like the local density of $Q_3$.

Starting from Eqs.~\eqref{eq:hn-intro} and \eqref{eq:bond-energy}, the Heisenberg equation gives
\begin{align}
i[H,e_n]
&=
i\left(\frac{J}{4}\right)^2\sum_m[h_m,h_n]\nonumber\\
&=
i\left(\frac{J}{4}\right)^2\Bigl([h_{n-1},h_n]+[h_{n+1},h_n]\Bigr)\nonumber\\
&=
i\left(\frac{J}{4}\right)^2\Bigl([h_{n-1},h_n]-[h_n,h_{n+1}]\Bigr).
\label{eq:appB-Heisenberg-step}
\end{align}
Comparing Eq.~\eqref{eq:appB-Heisenberg-step} with the continuity equation \eqref{eq:continuity}, one immediately obtains the local current quoted in Sec.~\ref{sec:II},
\begin{align}
j_n^E=
i\left(\frac{J}{4}\right)^2[h_{n-1},h_n],
\label{eq:appB-current-corrected}
\end{align}
which is identical to Eq.~\eqref{eq:local-current}. Summing over the chain and comparing with Eq.~\eqref{eq:Q3-intro-def}, one finds
\begin{align}
\mathcal{J}_E
=
-\left(\frac{J}{4}\right)^2 Q_3
=
-\frac{J}{4}\,i[\mathcal{B},H],
\label{eq:appB-total-current}
\end{align}
in agreement with Eq.~\eqref{eq:JE-Q3}. Thus the third conserved quantity is, up to the normalization fixed by Eq.~\eqref{eq:hn-intro} and the overall choice of current orientation, the physical energy current of the BT chain.




\section{Thermodynamic Bethe Ansatz for the modified spin-1 Babujian-Takhtajan model}
\label{app:C}
In this appendix, we derive the TBA used in Sec.~\ref{sec:III}. We follow the thermodynamic framework  of~\cite{Babuj-1983} for the BT model, and incorporate the third conserved charge through the single-particle energy derived in Appendix~\ref{app:B}. Throughout this work, we consider the antiferromagnetic regime $J>0$.

We start from the generalized Bethe-Ansatz equations for $s=1$~\cite{Fad-96},
\begin{align}
\label{BAE-1}
\left(\frac{\lambda_j+i}{\lambda_j-i}\right)^N
=
\prod_{\substack{k=1\\k\neq j}}^M
\frac{\lambda_j-\lambda_k+i}{\lambda_j-\lambda_k-i},
\quad j=1,\dots,M
\;,
\end{align}
where $\lambda_j$ are the rapidities, $M$ is the number of Bethe roots, and $N$ is the number of sites. The energy of the modified BT model, that can be derived via algebraic Bethe ansatz from the logarithmic derivative of the transfer matrix, is
\begin{align}
\label{Energy}
E_\alpha=-\sum_{j=1}^M\left(J-8\alpha\,\partial_j\right)\frac{1}{1+\lambda_j^2}.
\end{align}

In the thermodynamic limit the rapidities organize into Bethe strings,
\begin{align}
\label{Bethestrings}
\lambda_j^{(n,r)}=\lambda_j^n+i\left(\frac{n+1}{2}-r\right)
\;,
\quad r=1,\dots,n
\;,
\end{align}
where $\lambda_j^n$ is the real center of the $n$-string. If $M_n$ denotes the number of $n$-strings, then $\sum_{n=1}^\infty nM_n=M$. Multiplying Eq.~\eqref{BAE-1} over the string components gives
\begin{align}
\label{BAE-2}
e_n^N(\lambda_j^n)
=
\prod_{m=1}^{\infty}\prod_{k=1}^{M_m}E_{nm}(\lambda_j^n-\lambda_k^m)
\;,
\end{align}
with
\begin{align}
&e_n(x)=\frac{x+\frac{i n}{2}}{x-\frac{i n}{2}}
\;,\\
\label{Enm}
&E_{nm}(x)
=
e_{|n-m|}(x)e_{|n-m|+2}^2(x)\cdots e_{n+m-2}^2(x)e_{n+m}(x)\;.
\end{align}
Taking the logarithm yields
\begin{align}
\label{logBAE}
N\theta_{n,2}(\lambda_j^n)
=
2\pi I_j^n+
\sum_{m=1}^{\infty}\sum_{k=1}^{M_m}\Theta_{nm}(\lambda_j^n-\lambda_k^m)
\;,
\end{align}
where
\begin{align}
&\Theta_{nm}(x)=(1-\delta_{nm})\theta_{|n-m|}(x)
+2\theta_{|n-m|+2}(x)+\cdots\nonumber\\
&+2\theta_{n+m-2}(x)+\theta_{n+m}(x)
\;,
\end{align}
and
\begin{align}
&\theta_{n,p}(\lambda)
=
\sum_{l=1}^{\min(n,p)}\theta_{n+1+p-2l}(\lambda)\;,
\nn\\
&\theta_n(\lambda)=2\arctan\!\left(\frac{2\lambda}{n}\right)
\;.
\end{align}
For $s=1$, the left-hand side of Eq.~\eqref{logBAE} becomes
\begin{align}
\theta_{n,2}(\lambda)
=
\begin{cases}
\theta_{n+1}(\lambda)+\theta_{n-1}(\lambda)\;, & n\ge 2\;,\\[4pt]
\theta_2(\lambda)\;, & n=1
\;.
\end{cases}
\end{align}

The occupied $n$-strings are labeled by the Bethe quantum numbers $I_j^n$. The corresponding holes are labeled by $\tilde I_j^n$ and rapidities $\tilde\lambda_j^n$, satisfying the analogous equation. Introducing
\begin{align}
\label{hfunc}
2\pi h_n(\lambda)
=
\theta_{n,2}(\lambda)
-\frac{1}{N}\sum_{m=1}^{\infty}\sum_{k=1}^{M_m}
\Theta_{nm}(\lambda-\lambda_k^m)
\;,
\end{align}
one has $I_j^n=Nh_n(\lambda_j^n)$ and $\tilde I_j^n=Nh_n(\tilde\lambda_j^n)$. In the thermodynamic limit the particle and hole densities satisfy
\begin{align}
\rho_n(\lambda)+\rho_n^h(\lambda)=\frac{d h_n(\lambda)}{d\lambda}
\;.
\end{align}
Differentiating Eq.~\eqref{hfunc} gives
\begin{align}
&\rho_n(\lambda)+\rho_n^h(\lambda)
=
\frac{1}{2\pi}\frac{d\theta_{n,2}}{d\lambda}\\
&-\sum_{m=1}^{\infty}\int_{-\infty}^{\infty}
\frac{1}{2\pi}\frac{d\Theta_{nm}(\lambda-\lambda')}{d\lambda}\,\rho_m(\lambda')\,d\lambda'.\nonumber
\end{align}
Equivalently,
\begin{align}
\label{BAE-3}
\rho_n^h(\lambda)+\sum_{m=1}^{\infty}(A_{nm}*\rho_m)(\lambda)
=
\frac{\theta_{n,2}'(\lambda)}{2\pi}\;,
\quad n=1,2,\dots
\;,
\end{align}
where $*$ denotes convolution and
\begin{align}
A_{nm}
&=
\delta(\lambda)\delta_{nm}
+\frac{1}{2\pi}\frac{d\Theta_{nm}(\lambda-\lambda')}{d\lambda}\nonumber\\
&=
\delta(\lambda)\delta_{nm}
+(1-\delta_{nm})a_{|n-m|}(\lambda)
+2a_{|n-m|+2}(\lambda)\nonumber\\
&+\cdots
+2a_{n+m-2}(\lambda)+a_{n+m}(\lambda)
\;,
\end{align}
here $a_n(\lambda)$ is defined in Eq.~\eqref{eq:sec3-an}. The energy density and entropy density are given by
\begin{align}
\frac{E}{N}
&=
-\frac{1}{2}\sum_{n=1}^{\infty}\int_{-\infty}^{\infty}
\left(J-8\alpha\,\partial_\lambda\right)\theta_{n,2}'(\lambda)\,
\rho_n(\lambda)\,d\lambda
\;,\\
\frac{S}{N}
&=
\sum_{n=1}^{\infty}\int_{-\infty}^{\infty}
\Big[(\rho_n+\rho_n^h)\ln(\rho_n+\rho_n^h)-\rho_n\ln\rho_n \nonumber\\
&\hspace{1.2cm}
-\rho_n^h\ln\rho_n^h\Big]d\lambda
\;.
\end{align}
The equilibrium condition $\delta F=0$ for $F=E-TS$ gives
\begin{align}
\label{TBA-2}
&\ln\;\!\bigl(1+\eta_n(\lambda)\bigr)
=
-\frac{1}{2T}\left(J-8\alpha\,\partial_\lambda\right)\theta_{n,2}'(\lambda)\nonumber\\
&\hspace{1.2cm}
+\sum_{m=1}^{\infty}A_{nm}*\ln\,\!\bigl(1+\eta_m^{-1}(\lambda)\bigr)
\;,
\end{align}
where $\eta_n(\lambda)=\rho_n^h(\lambda)/\rho_n(\lambda)$. Applying $A^{-1}$ and using the identities displayed above reproduces the finite-temperature hierarchy quoted in Eq.~\eqref{eq:sec3-tba}. The corresponding free-energy density is
\begin{align}
\label{free-1}
f
&=
-\frac{T}{2\pi}\sum_{n=1}^{\infty}
\int_{-\infty}^{\infty}\theta_{n,2}'(\lambda)\ln\!\left(1+\eta_n^{-1}(\lambda)\right)d\lambda 
\;,\nonumber\\
&=
-\frac{1}{2}\int_{-\infty}^{\infty}
s(\lambda)\left(J-8\alpha\,\partial_\lambda\right)\theta_{2,2}'(\lambda)\,d\lambda \nonumber\\
&\hspace{1.2cm}
-T\int_{-\infty}^{\infty}s(\lambda)\ln\,\!\bigl(1+\eta_2(\lambda)\bigr)d\lambda
\;,
\end{align}
with $s(\lambda)$ defined in Eq.~\eqref{eq:sec3-skernel}.

\subsection{Low-temperature limit and flat phase}

Taking $T\to0$ in Eq.~\eqref{eq:sec3-tba} and using
\begin{align}
\lim_{T\to0}T\ln\!\left(1+e^{x/T}\right)=x^+\equiv\max(x,0)
\end{align}
we reproduce the zero-temperature dressed-energy equation given in Eq.~\eqref{eq:sec3-eps2full}, together with the definitions in Eqs.~\eqref{eq:sec3-gR} and \eqref{eq:sec3-gexplicit}. The free-energy density~\eqref{free-1} reduces to
\begin{align}
\label{free-low-T}
f
=
-J-\int_{-\infty}^{\infty}s(\lambda)\,\varepsilon_2^+(\lambda)\,d\lambda.
\end{align}
For $\alpha=0$, this recovers the BT ground-state energy, $f=-J$. We next establish the flat phase stated in Eq.~\eqref{eq:sec3-flat-phase}. For $\alpha/J\le 1/(8\pi)$, Eq.~\eqref{eq:sec3-gexplicit} gives $g(\lambda)\le0$ for all $\lambda$. Moreover, by working in Fourier space,
\begin{align}
\widehat s(\omega)=\frac{1}{2\cosh(\omega/2)},
\qquad
\widehat a_n(\omega)=e^{-n|\omega|/2}
\;,
\end{align}
so that
\begin{align}
\widehat R(\omega)=\widehat s(\omega)^2+\widehat s(\omega)\widehat a_1(\omega)
\;,
\quad
\|R\|_1=\widehat R(0)=\frac{3}{4}
\;.
\end{align}
Since $R(\lambda)\ge0$, Eq.~\eqref{eq:sec3-eps2full} implies
\begin{align}
0\le \varepsilon_2^+(\lambda)
\le
\int_{-\infty}^{\infty}R(\lambda-\mu)\,\varepsilon_2^+(\mu)\,d\mu,
\end{align}
and therefore
\begin{align}
\|\varepsilon_2^+\|_\infty
\le
\|R\|_1\,\|\varepsilon_2^+\|_\infty
=
\frac{3}{4}\,\|\varepsilon_2^+\|_\infty.
\end{align}
The only possibility is $\varepsilon_2^+(\lambda)\equiv0$, which proves Eq.~\eqref{eq:sec3-flat-phase}. In other words, below threshold the $Q_3$ deformation tilts the source but never empties any part of the two-string sea, so the ground-state energy remains pinned at its undeformed value.

\subsection{Two-boundary solution above the transition}

For $\alpha>\alpha_c$, Eq.~\eqref{eq:sec3-sign} shows that $\varepsilon_2^+(\lambda)$ coincides with $\varepsilon_2(\lambda)$ outside the interval $[b_-,b_+]$ and vanishes inside it. Equation~\eqref{eq:sec3-eps2full} can therefore be rewritten on the full line as
\begin{align}
(1-R)*\varepsilon_2
=
g-\int_{b_-}^{b_+}R(\lambda-\mu)\,\varepsilon_2(\mu)\,d\mu
\;.
\label{eq:C-rewrite}
\end{align}
This is the crucial step: the finite-interval equation is obtained by projecting the exact full-line problem, rather than imposing a one-boundary ansatz.

The inverse of $1-R$ is most conveniently evaluated in Fourier space. Using the transforms above,
\begin{align}
1-\widehat R(\omega)=\frac{1}{(1+e^{-|\omega|})^2}
\;,
\label{eq:one-minus-R}
\end{align}
from which it follows that
\begin{align}
&\frac{\widehat s(\omega)}{1-\widehat R(\omega)}=e^{-|\omega|/2}+e^{-3|\omega|/2} \;,
\nonumber\\ 
&\frac{\widehat R(\omega)}{1-\widehat R(\omega)}=2e^{-|\omega|}+e^{-2|\omega|}
\;.
\end{align}
These expressions are precisely the Fourier transforms of the functions $q(\lambda)$ and $K(\lambda)$ introduced in Eq.~\eqref{eq:sec3-qK}. Acting with $(1-R)^{-1}$ on Eq.~\eqref{eq:C-rewrite} gives
\begin{align}
&\varepsilon_2(\lambda)
=\\
&-\pi\left(J-8\alpha\,\partial_\lambda\right)q(\lambda)
-\int_{b_-}^{b_+}K(\lambda-\mu)\,\varepsilon_2(\mu)\,d\mu
\nn \;,
\label{eq:C-reconstruct}
\end{align}
valid for all $\lambda$. Restricting Eq.~\eqref{eq:C-reconstruct} to $\lambda\in[b_-,b_+]$ reproduces Eq.~\eqref{eq:sec3-two-boundary}, while the endpoints are fixed by Eq.~\eqref{eq:sec3-boundaries}. This also clarifies the inconsistency of the one-boundary truncation: once the postcritical support of $\varepsilon_2^+$ is treated on the full line, the physical solution closes on one connected occupied interval and therefore exhibits two self-consistent edges.

The free-energy formula quoted in Eq.~\eqref{eq:sec3-free} follows from the same identity. Multiplying Eq.~\eqref{eq:C-rewrite} by $q(\lambda)$ and integrating over $\lambda$ gives
\begin{align}
\int_{-\infty}^{\infty}q(\lambda)g(\lambda)\,d\lambda
&=
\int_{-\infty}^{\infty}s(\lambda)\varepsilon_2(\lambda)\,d\lambda \nonumber\\
&\hspace{0.5cm}
+\int_{b_-}^{b_+}\varepsilon_2(\mu)\!
\int_{-\infty}^{\infty}q(\lambda)R(\lambda-\mu)\,d\lambda\,d\mu
\;.
\end{align}
Using $(1-R)*q=s$, or equivalently $q*R=q-s$, this becomes
\begin{align}
\int_{-\infty}^{\infty}q(\lambda)g(\lambda)\,d\lambda
&=
\int_{b_-}^{b_+}q(\lambda)\varepsilon_2(\lambda)\,d\lambda \nn\\
&\hspace{0.5cm}
+\int_{-\infty}^{\infty}s(\lambda)\varepsilon_2^+(\lambda)\,d\lambda
\;.
\label{eq:C-qg-identity}
\end{align}
The left-hand side is $-J$: the term proportional to $q(\lambda)s'(\lambda)$ is odd and integrates to zero, while the remaining piece gives $-\pi J\int q(\lambda)s(\lambda)\,d\lambda=-J$. Combining Eq.~\eqref{eq:C-qg-identity} with Eq.~\eqref{free-low-T} yields Eq.~\eqref{eq:sec3-free}.

Finally, differentiating Eq.~\eqref{eq:sec3-two-boundary} with respect to $\alpha/J$ produces no boundary terms because the derivatives of the limits are multiplied by $\varepsilon_2(b_\pm)=0$. The Hellmann--Feynman relation quoted in Eq.~\eqref{eq:q3-derivative-intro} therefore remains exact in the two-boundary regime.

\subsection{Expansion near \texorpdfstring{$\alpha_c$}{TEXT}}
\label{app:alphac-expansion}

Because Eq.~\eqref{eq:sec3-flat-phase} is exact below threshold, the expansion around $\alpha_c$ is necessarily one-sided. The instability first appears in the far-left rapidity tail of Eq.~\eqref{eq:sec3-gexplicit}, so the leading singular term is controlled by the left boundary $b_-$. The right boundary is generated only through backflow and contributes to Eq.~\eqref{free-low-T} only at higher order.

Writing $\lambda=b_- - x$ with $x\ge0$ and expanding Eq.~\eqref{eq:sec3-gexplicit} at fixed $x$, one obtains that the left-tail source is of order $(8\pi\alpha/J-1)^{3/2}$. It is therefore natural to define the edge profile by
\begin{align}
\label{eq:C-edge-profile}
\frac{\varepsilon_2(b_- - x)}{J}
&=
\frac{\pi}{\sqrt{2}}
\left(8\pi\frac{\alpha}{J}-1\right)^{3/2}\phi(x)\\
&+O\!\left[\left(\frac{\alpha}{J}-\frac{1}{8\pi}\right)^{5/2}\right]
\;.
\nn
\end{align}
Here, the exact leading edge profile $\phi(x)$ is given by $\phi(x)=e^{\pi c_0}\phi_{\pi}(x)-e^{3\pi c_0}\phi_{3\pi}(x)$, where
$\phi_s(x)$ denotes the solution of
\begin{equation}
\phi_s(x)-\int_0^{\infty}R(x-y)\phi_s(y)\,dy=e^{-s x}\,,\quad x>0.
\end{equation}

Solving the resulting half-line Wiener-Hopf problem and imposing $\varepsilon_2(b_-)=0$ gives two exact outputs. The first is the asymptotic position of the left boundary,
\begin{align}
\label{eq:C-bminus-asymptotic}
b_-
&=
\frac{1}{2\pi}\log\!\left(\frac{8\pi\,\alpha/J-1}{8\pi\,\alpha/J+1}\right)+\frac{1}{2\pi}\log\!\left(\frac{4e^2}{27}\right)\\
&+O\!\left(\frac{\alpha}{J}-\frac{1}{8\pi}\right) 
\;.
\nn
\end{align}
The second is that the normalized profile satisfies
\begin{align}
&\phi(x)-\int_0^{\infty}R(x-y)\phi(y)\,dy\nonumber
\\
&= \left(\frac{4e^2}{27}\right)^{1/2}e^{-\pi x}
 - \left(\frac{4e^2}{27}\right)^{3/2}e^{-3\pi x},
\label{eq:C-edge-WH}
\\
&\left(\frac{4e^2}{27}\right)^{1/2}
\int_0^{\infty}e^{-\pi x}\phi(x)\,dx
= \frac{\pi}{27}
\label{eq:C-edge-moment}
\;.
\end{align}
Thus, the constant $c_0$ in the normalized profile is fixed by
\begin{equation}
e^{\pi c_0}=\frac{2e}{\sqrt{27}}.
\end{equation}
Substituting Eqs.~\eqref{eq:C-edge-profile} and \eqref{eq:C-edge-moment} into Eq.~\eqref{free-low-T} gives
\begin{align}
\frac{f}{J}+1
&=
-\frac{\pi}{2}
\left(8\pi\frac{\alpha}{J}-1\right)^2
\left(\frac{4e^2}{27}\right)^{1/2}
\int_0^{\infty}e^{-\pi x}\phi(x)\,dx
\nonumber\\
&\hspace{0.6cm}
+O\!\left[\left(\frac{\alpha}{J}-\frac{1}{8\pi}\right)^3\right]
\nonumber\\
&=
-\frac{\pi^2}{54}
\left(8\pi\frac{\alpha}{J}-1\right)^2
+O\!\left[\left(\frac{\alpha}{J}-\frac{1}{8\pi}\right)^3\right].
\end{align}
Equivalently,
\begin{align}
\frac{f}{J}
=
-1
-\frac{32\pi^4}{27}
\left(\frac{\alpha}{J}-\frac{1}{8\pi}\right)^2
+O\!\left[\left(\frac{\alpha}{J}-\frac{1}{8\pi}\right)^3\right].
\label{eq:C-f-series}
\end{align}
Differentiating Eq.~\eqref{eq:C-f-series} and using Eq.~\eqref{eq:q3-derivative-intro} gives the exact one-sided onset of the conserved charge density,
\begin{align}
\frac{1}{N}\langle Q_3\rangle
=
-\frac{64\pi^4}{27}
\left(\frac{\alpha}{J}-\frac{1}{8\pi}\right)
+O\!\left[\left(\frac{\alpha}{J}-\frac{1}{8\pi}\right)^2\right].
\label{eq:C-q3-series}
\end{align}
In dimensionful form,
\begin{align}
\frac{1}{N}\langle Q_3\rangle
=
-\frac{64\pi^4}{27J}(\alpha-\alpha_c)
+O\!\left[\frac{(\alpha-\alpha_c)}{J}\right]^2
\;.
\label{eq:C-q3-series-dimful}
\end{align}
Thus, the free energy starts quadratically above threshold, while the conserved current turns on linearly with an exactly determined slope. The coefficient is already renormalized at leading order by the self-consistent Wiener--Hopf dressing of the left edge.


\end{document}